\documentclass[]{imsart}

\usepackage[]{graphicx}\usepackage[]{color}

\makeatletter
\def\maxwidth{ %
  \ifdim\Gin@nat@width>\linewidth
    \linewidth
  \else
    \Gin@nat@width
  \fi
}
\makeatother

\definecolor{fgcolor}{rgb}{0.345, 0.345, 0.345}

\usepackage{framed}
\makeatletter
 {\par\unskip\endMakeFramed%
 \at@end@of@kframe}
\makeatother

\definecolor{shadecolor}{rgb}{.97, .97, .97}
\definecolor{messagecolor}{rgb}{0, 0, 0}
\definecolor{warningcolor}{rgb}{1, 0, 1}
\definecolor{errorcolor}{rgb}{1, 0, 0}

\usepackage{alltt} 
\usepackage[utf8]{inputenc}
\usepackage{todonotes}
\usepackage{latexsym}
\usepackage{amssymb}
\usepackage{epsfig}
\usepackage{amsmath}
\usepackage{xcolor}
\usepackage{float}
\usepackage{hyperref}
\usepackage{subfig}
\usepackage{multirow}
\usepackage{booktabs}
\usepackage[title]{appendix}
\usepackage{algorithm} 
\usepackage{algpseudocode} 
\makeatletter
\newcommand{\algmargin}{\the\ALG@thistlm}
\makeatother
\newlength{\whilewidth}
\algnewcommand{\parState}[1]{\State%
	\parbox[t]{\dimexpr\linewidth-\algmargin}{\strut #1\strut}}

\IfFileExists{upquote.sty}{\usepackage{upquote}}{}

\usepackage{enumitem}
\usepackage{orcidlink}
\usepackage{todonotes}
\RequirePackage[OT1]{fontenc}
\RequirePackage{amsmath,amssymb,amsthm}
\RequirePackage[round]{natbib}

\hypersetup{
  citecolor=black}
	
\RequirePackage[colorlinks,citecolor=blue,urlcolor=blue]{hyperref}
\usepackage{graphicx}
\usepackage{mathtools}

\startlocaldefs
\numberwithin{equation}{section}
\theoremstyle{plain}

\endlocaldefs

\theoremstyle{plain}
\long\def\comment#1{}

\theoremstyle{definition}

\numberwithin{definition}{section}

\numberwithin{remark}{section}

\renewcommand{\qed}{\hfill{\tiny \ensuremath{\blacksquare} }}%

\makeatletter
\newcommand{\setword}[2]{%
  \phantomsection
  #1\def\@currentlabel{\unexpanded{#1}}\label{#2}%
}
\makeatother

\begin{document}

\begin{frontmatter}

\title{Sensitivity Analysis for Causal ML: A Use Case at Booking.com}
\runtitle{Sensitivity Analysis for Causal ML at Booking.com}
\runauthor{Bach et al.}
\thankstext{T1}{Version: October, 2025. Corresponding author: philipp.bach@fu-berlin.de}

\author{%
Philipp Bach\orcidlink{0000-0002-7183-9239}\thanksref{T1},
Victor Chernozhukov\orcidlink{0000-0002-3250-6714},
Carlos Cinelli\orcidlink{0000-0002-2021-7739},
Lin Jia\orcidlink{0009-0000-7819-670X},
Nils Skotara\orcidlink{0009-0008-4763-6555},
Sven Klaassen\orcidlink{0009-0004-9080-0809},
Martin Spindler\orcidlink{0000-0002-1294-7782}
\thanks{%
Affiliations:%
~Philipp Bach -- Free University of Berlin; %
Victor Chernozhukov -- MIT; %
Carlos Cinelli -- University of Washington; %
Lin Jia and Nils Skotara -- Booking.com; %
Sven Klaassen and Martin Spindler -- University of Hamburg and Economic AI.
}
}

\begin{abstract}
Causal Machine Learning has emerged as a powerful tool for flexibly estimating causal effects from observational data in both industry and academia. However, causal inference from observational data relies on untestable assumptions about the data-generating process, such as the absence of unobserved confounders. When these assumptions are violated, causal effect estimates may become biased, undermining the validity of research findings. In these contexts, sensitivity analysis plays a crucial role, by enabling data scientists to assess the robustness of their findings to plausible violations of unconfoundedness. This paper introduces sensitivity analysis and demonstrates its practical relevance through a (simulated) data example based on a use case at Booking.com. We focus our presentation on a recently proposed method by \citet{chernozhukov2023long}, which derives general non-parametric bounds on biases due to omitted variables, and is fully compatible with (though not limited to) modern inferential tools of Causal Machine Learning. By presenting this use case, we aim to raise awareness of sensitivity analysis and highlight its importance in real-world scenarios.
\end{abstract}

\begin{keyword}
\kwd{Causal Machine Learning}
\kwd{Sensitivity Analysis}
\kwd{Robustness}
\kwd{Unobserved Confounding}
\kwd{Omitted Variable Bias}
\kwd{Double/Debiased Machine Learning}
\end{keyword}

\end{frontmatter}

\section{Introduction}

Randomized controlled trials (RCTs) (also commonly known in industry as ``A/B tests,'' when the treatment consists of two categories) are widely regarded as the gold standard for estimating  causal effects. In an RCT, participants are randomly assigned to either a control group (A) or a treatment group (B), thus preventing individuals from self-selecting into a treatment status. This randomization allows for clear attribution of any differences in outcomes between groups directly to the treatment itself, rather than to other confounding factors.  However, due to constraints such as high costs, ethical concerns, or practical limitations, conducting such experiments is often infeasible. In these situations, data analysts may then turn to observational data to infer causal effects.
 
Unlike experimental studies,  observational studies are based on historically collected data, such as past transactions, with no experimental control over the treatment assignment process.  In these studies, the treatment and outcome may be associated even without a true causal relationship. This association can occur because individuals' treatment choices and outcomes may both be influenced by confounding factors.  To draw valid causal inferences, it is essential to account for these factors effectively. One widely used approach to address these concerns is to adjust for observed covariates, aiming to mitigate confounding biases. This can be done flexibly and efficiently with modern tools of Causal Machine Learning. The validity of this approach, however, still relies on the assumption of unconfoundedness, also known as ``selection-on-observables'' \citep{angrist2009mostly}. This assumption posits that, conditional on the observed covariates, the treatment can be considered \textit{as good as randomly assigned}. Importantly, it assumes no unobserved confounding remains. What if this is not true? How biased would our estimates be?


To illustrate these concepts, we will explore a practical simulated example inspired by a real use case at Booking.com.\footnote{We note that the use case is presented in a stylized way. Moreover, due to confidentiality concerns, and for the sake of reproducibility, the empirical analysis is based on a simulated data set. The results of this simulation cannot be used to recover the findings from the actual use case at Booking.com.}  Booking.com is one of the world's leading digital travel platforms and offers not only accommodations, but also other products such as flights, rental cars, and other travel-related services. In this application, we are concerned with estimating the causal effect of cross-selling products on customer relationship. In general, cross-selling, i.e., selling additional products to customers that purchase one product, is an important marketing strategy to increase revenue, customer satisfaction and customer loyalty in the long term. In our analysis, the treatment of interest is whether a customer books an ancillary product in addition to their accommodation booking. The outcome measures the number of follow-up accommodation bookings made within six months after the initial booking. Note an A/B test, which could provide direct experimental evidence of the effect of the treatment on the outcome, is not feasible in this context due to high costs and ethical concerns associated with forcing customers to accept products they may not be interested in.\footnote{Other research designs, such as encouragement designs, could be considered.}

Given the constraints of not being able to conduct a randomized controlled trial, we rely on observational data, where the treatment assignment stems from customers' choices rather than random allocation. This approach introduces potential confounding factors, as customers’ decisions to book ancillary products are influenced by various factors that may also impact their follow-up bookings. For example, the purpose of a trip — such as whether it is for business or leisure — can significantly affect the likelihood of purchasing additional services like a taxi transfer. Business travelers may be more inclined to book such services compared to leisure travelers. However, these same factors also influence the number of follow-up accommodation bookings. To draw valid causal inferences from observational data, it is thus crucial to account for these confounding variables to isolate the true effect of the treatment. But what if some confounding factors were not measured?

A key concern in this application is whether the observed variables are sufficient to account for all relevant confounding factors, or if there are any unobserved variables that could bias the results. For instance, prior customer loyalty might be a significant factor influencing both the decision to purchase ancillary products and the frequency of follow-up bookings. If this factor is not adequately captured in the data, the assumption of unconfoundedness --- where treatment is as good as randomly assigned given observed covariates --- might be violated. Sensitivity analysis can then be used to assess how robust our  conclusions are to such violations. In some cases, it may suggest that causal inferences are still warranted, even if we did not account perfectly for all confounding factors, such as customer loyalty. Our goal in this paper is to introduce the concepts of causal effect identification, estimation and sensitivity analysis, using the flexible tools of Causal Machine Learning, through this practical example.

The rest of the paper is organized as follows. Section~\ref{sec:obsci}, briefly reviews identification via covariate adjustment, and estimation using Debiased Machine Learning (DML) \citep{chernozhukov2018double}. Section~\ref{sec:sensitivity}, motivates sensitivity analysis, and reviews the recent proposal of \citet{chernozhukov2023long} for sensitivity analysis in Causal Machine Learning. Section~\ref{sec:usecase} demonstrates this approach in the use case from Booking.com. We complement our analysis with a reproducible notebook\footnote{\url{https://docs.doubleml.org/stable/examples/py_double_ml_sensitivity_booking.html}.}, and describe the implementation using the \texttt{DoubleML} Python library in Section \ref{implementation} of the Supplementary Material.

\section{Observational Causal Inference} \label{sec:obsci}

\subsection{Potential outcomes and causal parameters}

Causal inference starts by \emph{defining} the causal effects of interest, which measure how experimental manipulations of a treatment \(D\) would influence an outcome \(Y\). These effects are often formulated in terms of \textit{potential outcomes}. For instance, consider a binary treatment variable, where $D=1$ denotes treatment, and $D=0$ denotes the control condition. Here, we write $Y(1)$ to denote the potential outcome of an individual if, perhaps contrary to fact, they were assigned to the treatment condition; analogously, $Y(0)$  denotes the potential outcome of an individual if, perhaps contrary to fact, they were assigned to the control condition. 

The Average Treatment Effect (ATE) measures the average difference in expected outcomes if the treatment were applied to the entire population compared to this average if the treatment were instead withheld:
\[
\theta = \mathbb{E}[Y(1)]- \mathbb{E}[Y(0)].
\]
Similarly, the Average Treatment effect on the Treated (ATT) measures the average treatment effect specifically for those individuals who actually receive the treatment:
\[
\theta_0 = \mathbb{E}[Y(1) \mid D=1]- \mathbb{E}[Y(0) \mid D=1].
\]

Here we will focus on the ATT, as it is the key causal parameter for our business use case at Booking.com. The ATT measures the average impact on follow-up bookings that results from booking an ancillary product, specifically among those who have actually made such a booking. If the treatment effect varies across individuals and individuals self-select into the treatment, the ATT may differ from the ATE.

Note, however, that we do not have any data on the \emph{potential outcomes} under different experimental conditions. We have only observational data on the actual outcome $Y$, treatment $D$ and observed covariates $X$. In order to estimate the ATT, we need assumptions that connect the observational data with the counterfactual parameter we are interested in. The task of deciding whether a set of assumptions is sufficient to recover the causal parameter of interest is often referred to as an \emph{identification problem}.

\subsection{Identification under unconfoundedness}

The ATT  can be \textit{identified} from the distribution of observed data $W_s = (Y, D, X)$  under the following assumptions:
\begin{enumerate}

\item \textit{Unconfoundedness}: $\{Y(0), Y(1)\} \perp\!\!\!\perp D \vert X$,

\item \textit{Overlap}: $0 < P(D=1 \vert X) < 1$,

\item \textit{Consistency}: $Y=Y(D)$.

\end{enumerate}
Unconfoundedness states that given observed covariates $X$, the treatment is independent of potential outcomes, implying no unobserved selection mechanisms. Overlap ensures a non-zero probability of receiving treatment for all covariate values. Consistency connects observed outcomes with potential outcomes, and implicitly assumes a well-defined treatment without spillover effects or hidden variations in the treatment.\footnote{This is sometimes also called SUTVA---the Stable Unit Treatment Value Assumption.}  The derivation goes as follows:
\begin{align}
\theta_0 &= \mathbb{E}[Y(1) \mid D=1]- \mathbb{E}[Y(0) \mid D=1]\\
&= \mathbb{E}[Y(1) \mid D=1]- \mathbb{E}[\mathbb{E}[Y(0) \mid D=1, X]|D=1]\\
&= \mathbb{E}[Y(1) \mid D=1]- \mathbb{E}[\mathbb{E}[Y(0) \mid D=0, X]|D=1]\\
&= \mathbb{E}[Y \mid D=1]- \mathbb{E}[\mathbb{E}[Y \mid D=0, X]|D=1] 
\end{align}
The first line restates the definition of the ATT. The second line uses the law of total probability (and overlap to avoid division by zero). The third line applies the unconfoundedness assumption, and the fourth line uses the consistency assumption.

We have thus established that, given the stated assumptions, the counterfactual parameter \(\theta_0\) can be expressed as a population quantity based solely on observational data, specifically:
\[
\theta_0 = \mathbb{E}[Y \mid D=1] - \mathbb{E}[\mathbb{E}[Y \mid D=0, X]|D=1].
\]
In practice, however, we do not have access to population quantities. Therefore, we must estimate $\theta_0$ from finite samples.

\begin{figure}[t]
\includegraphics[width=0.60\linewidth]{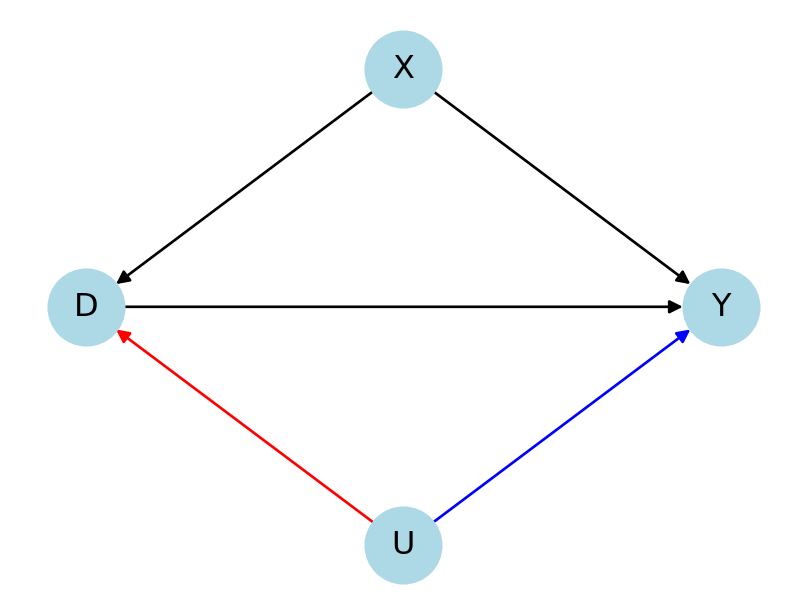}
\caption{A stylized DAG illustrating an observed and unobserved confounding relationship.}
\label{fig:dag1}
\end{figure} 

\subsection{Estimation with Causal ML}
\label{sec:DML}

Various estimation procedures can be used for estimating causal effects from data, including matching, regression, and propensity-score-based methods. Here we focus on causal estimation using machine learning (ML) algorithms. While ML methods can be powerful, their naive application may introduce biases in the estimation procedure that prevent proper uncertainty quantification and statistical inference. 

To illustrate this point, let us rewrite the ATT as:
$$
\theta_0 = \mathbb{E}[Y|D=1]  -\mathbb{E}[g_s(0,X)|D=1],
$$ 
where here $g_s(D, X):= E[Y|D, X]$ is the conditional expectation of $Y$ given $D$ and $X$. The first component of the ATT, namely, $\mathbb{E}[Y|D=1]$, can be easily estimated from data with the sample average of the outcome in the treated group. The second component of the ATT, on the other hand,  can be harder to estimate. The function $g_s(D,X)$ can be complex or high-dimensional, motivating the use of ML algorithms. A natural estimator for this quantity would be to directly employ ML estimators for $g_s(D,X)$ and estimate the component $\mathbb{E}[g_s(0,X)|D=1],$ of the ATT by taking the empirical mean of $\hat{g}_s(0,X)$ in the subgroup of treated units. This, however, would lead to severe biases and slow convergence, with estimation errors that are not asymptotically normally distributed \citep{chernozhukov2018double,doubleml2024R,DoubleML2022Python}, precluding standard statistical inference.

Hence, ML-based causal estimation requires certain adjustments to work in practice. A leading framework for such adjustments is the Double/Debiased Machine Learning approach proposed \citet{chernozhukov2018double}, which is based on three key ingredients \citep{doubleml2024R}:\footnote{Here, we briefly sketch the building blocks of DML. For more a more formal introduction, we refer to \citet{chernozhukov2018double}. An intuitive introduction with multiple examples is available from \citet{doubleml2024R}.} 
\begin{enumerate}
\item Neyman Orthogonal Scores;
\item High-quality Machine Learners; and,
\item Sample Splitting.
\end{enumerate}

Neyman orthogonality alleviates the regularization bias problem by reducing the sensitivity of the target parameter to estimation errors of nuisance functions such as $g_s(D,X)$. Perhaps counter-intuitively, this often involves rewriting the target parameter as a function of two nuisance parameters instead of one. In the ATT example, for instance, it involves estimating the propensity score $m_s(X) := E[D|X]$, and estimating the target parameter with a combination of regression adjustment and inverse probability of treatment weighting. Neyman-orthogonal scores are often available from the literature --- several examples, including the ATT, are summarized in \citet{chernozhukov2018double}. 

The second key ingredient for DML to work is the quality of the machine learning algorithms used to estimate the nuisance parameters, in terms of convergence rates. In general, they need to have convergence rates faster than $n^{-1/4}$ to guarantee that the higher-order biases vanish.  Structural conditions for achieving these learning rates  are known for many ML estimators (e.g. the Lasso learner under sparsity). In practice this also demands the careful choice of learners and hyperparameters \citep{bach2024hyperparameter}.

The third ingredient, sample splitting, requires that the ML estimation of the nuisance parameters are fitted on a partition of the data set separate from the data used for calculating the target causal parameter. The role of the train and test samples can be swapped, which is called \textit{cross-fitting} --- the explicit DML algorithm is presented in Section \ref{dmlalgo} in the Supplementary Material.  \citet{chernozhukov2018double} then show that the DML causal estimator $\hat{\theta}_0$, built on these three ingredients,  concentrates around the true value $\theta_0$ and is asymptotically normal, thus enabling standard statistical inference. 

\section{Sensitivity Analysis} \label{sec:sensitivity}

\subsection{Background}

In the previous section, we discussed assumptions that allowed us to estimate the Average Treatment Effect on the Treated from observational data, specifically unconfoundedness, overlap, and consistency. However, in general, observed covariates \(X\) might not be sufficient to control for confounding due to the presence of unobserved confounders.  Figure \ref{fig:dag1} illustrates a stylized confounding scenario with both observed confounding \(X\) and unobserved confounding \(U\), which prevents an unbiased estimation of the causal parameter of interest.

When the unconfoundedness assumption is not fully met, sensitivity analysis provides a way to understand the potential impact of unobserved confounding. The goal is to quantify the influence of unobserved confounders through sensitivity parameters, which measure the strength of confounding in both the treatment assignment mechanism and the outcome equation (illustrated by the red and blue arrows in Figure~\ref{fig:dag1}). This usually involves varying  sensitivity parameters and assessing how different scenarios impact our causal effect estimates. This can be done through numerical simulations and recalculations of the causal effect estimate under different scenarios, or by positing parameter values into an analytical bias formula.

There is an immense literature on sensitivity analysis spanning the fields of statistics, econometrics, and computer science including the work by \cite{cornfield1959smoking,rosenbaum1983assessing, robins1999association, frank:smr2000, rosenbaum2002gamma,imbens2003sensitivity, brumback2004sensitivity,frank:eepa2008, hosman2010sensitivity, imai2010identification, arah2011, blackwell2013selection, frank2013would,carnegie:jree2016, dorie2016flexible, middleton2016bias, oster:jbes2017, cinelli:icml2019, cinelli2020making, cinelli2020sensemakr, zhang2021exploiting, cinelli2022omitted}. This body of work includes a variety of approaches, each with different assumptions regarding the strength and nature of unobserved confounding, as well as the specific causal parameters they address. Due to this diversity and complexity, a comprehensive and exhaustive survey of all methods is beyond the scope of this work. Here we focus on the omitted variable bias approach of \citet{chernozhukov2023long}. This approach is  nonparametric, while also permitting (semi-)parametric restrictions (such as partial linearity) when such assumptions are made.  It covers a broad range of common causal parameters often investigated in causal inference studies, such as averages of potential outcomes, average treatment effects, average causal derivatives, and policy effects resulting from covariate shifts. Notably, it covers the Average Treatment Effect on the Treated relevant to the demonstration use case. This approach also makes it possible to use modern tools of Causal ML for estimation and inference, as described in Section~\ref{sec:DML}.

\subsection{Omitted Variable Bias in Causal ML}

The omitted variable bias approach of \citet{chernozhukov2023long} starts by positing that unconfoundedness holds given access to the full data $W=\{Y, D, X, U\}$, including the unobserved confounders $U$, as in Figure~\ref{fig:dag1}. Had we had access to this data, we would be able to estimate the target causal parameter of interest (also called the \textit{long} parameter), i.e, the ATT corresponding to an adjustment by $X$ and $U$,
$$
\theta_0 = \mathbb{E}[Y|D=1] - \mathbb{E}[\mathbb{E}[Y|D=0, X, U]|D=1].
$$

However, since $U$ is unobserved, the data scientist has access to the observable data only, i.e., $W_s=\{Y, D, X\}$.  Similarly, she can only estimate the \emph{short} parameter $\theta_s$, which adjusts for $X$, but $U$ is \emph{omitted} from the regression equation, 
$$
\theta_s = \mathbb{E}[Y|D=1] - \mathbb{E}[\mathbb{E}[Y|D=0, X]|D=1].
$$
The goal is then to bound the bias due to the omission of $U$, that is, the difference in the long and short estimands:
\begin{align}
\text{bias} =\theta_s - \theta_0.
\end{align}
\citet{chernozhukov2023long} show that this bias can be expressed as,
\begin{align}
\text{bias}^2 = \rho^2 C_Y^2 C_D^2  S^2, \label{eq:bias}
\end{align}
where here $\rho^2$, $C_Y^2$ and $C_D^2$ are sensitivity parameters that must be constrained by hypothesis that limit the strength of unobserved confounding, and $S^2$ is a scaling factor which is estimable from the observed data $W_s$. In fact,  leveraging the Riesz representation theorem, \citet{chernozhukov2023long} show that the same bias formula holds more generally for any target parameter that can expressed as a linear functional of the conditional expectation function of the outcome. Here we focus on the ATT as it is the relevant parameter for the demonstration use case---see Section~\ref{sec:generalformulation} for the general formulation.

For the ATT, these sensitivity parameters have the following interpretation. The parameter \(\rho^2 \in [0,1]\) measures the correlation between the confounding errors in the outcome regression \(\mathbb{E}[Y \mid D, X, U]\) and the treatment regression \(\mathbb{E}[D \mid X, U]\). For an unobserved variable \(U\) to introduce bias in the ATT estimation, it is not enough for \(U\) to simply explain variations in both treatment and outcome; the errors in these regressions need to be systematically related. In the absence of additional assumptions about the data-generating process, this parameter is typically set to its upper bound of 1, though less conservative scenarios can also be considered.

The parameters \(C_Y^2\) and \(C_D^2\) quantify how confounding influences the outcome and treatment assignment mechanisms, respectively. They parameterize the blue and red arrows of Figure~\ref{fig:dag1} in terms of the relevant quantities for assessing the omitted variable bias of the ATT. Specifically, \(C_Y^2 \in [0,1]\) is defined as:
\[
C_Y^2 := \frac{\text{Var}(\mathbb{E}[Y \mid D, X, U]) - \text{Var}(\mathbb{E}[Y \mid D, X])}{\text{Var}(Y) - \text{Var}(\mathbb{E}[Y \mid D, X])} =: R^2_Y.
\]
This quantity, which we also denote by \(R^2_Y\), represents the (nonparametric) partial \(R^2\) of \(U\) with respect to \(Y\). In other words, it measures the proportion of the residual variation in the outcome \(Y\) that is explained by the unobserved confounder \(U\) after accounting for the variation explained by the observed covariates \(D\)  and \(X\). Notably, \(R^2_Y\) can be set to its upper bound of 1, and this still results in a finite bias. 

Finally, the parameter $C_D^2 \in [0, \infty)$ is given by,
\begin{align*}
C_D^2 = \frac{\mathbb{E}\left[O(X, U)\right] - \mathbb{E}\left[O(X)\right]}{\mathbb{E}\left[O(X)\right]},
\end{align*}
where \( O(X, U) := \frac{P(D=1 \mid X, U)}{1 - P(D=1 \mid X, U)} \) and \( O(X) := \frac{P(D=1 \mid X)}{1 - P(D=1 \mid X)} \) represent the odds of receiving treatment, conditional on \(\{X, U\}\) and \(X\), respectively. This quantifies the increase in the average odds of receiving treatment due to the presence of the unobserved confounder \(U\). In other words, \(C^2_D\) measures how much the unobserved confounder \(U\) improves our ability to predict, on average, whether individuals are in the treated or control group, compared to when we only have access to the observed confounders.

Since \(C_D^2\) is unbounded, it may be useful to re-express it as an \(R^2\)-like measure, similar to $R^2_Y$, for interpretability purposes:
\[
C_D^2 = \frac{R^2_D}{1 - R^2_D},
\]
where
\[
R^2_D := \frac{\mathbb{E}\left[O(X, U)\right] - \mathbb{E}\left[O(X)\right]}{\mathbb{E}\left[O(X, U)\right]} \in [0, 1].
\]
The $R^2_D$ parameter quantifies the proportion of  the average odds of receiving treatment explained by the unobserved confounder \(U\), after accounting for what is explained by $X$. Unlike the other parameters, which can be set to a worst-case value of 1, \(R^2_D\) must always be less than 1 to ensure that the bias remains finite. That is, in order to have informative bounds, we always need to constraint how much confounders affect the treatment assignment.

\paragraph{Estimation and inference} All results above hold for population parameters. In practice, we need to estimate the bounds on the target parameter from finite samples. \citet{chernozhukov2023long} show that estimation and inference for sensitivity analysis can be performed using the same principles of Double/Debiased Machine Learning, as briefly discussed in Section~\ref{sec:DML}.

\paragraph{Robustness values} The previous bias formula allow us to assess how much bias any confounding scenario, as given by specific values of sensitivity parameters $R^2_Y$ and $R^2_D$, would cause. Sometimes, however, researchers may not have a particular scenario in mind. In such cases, users can still report \emph{robustness values} \citep{cinelli2020making, chernozhukov2023long}, summarizing the minimal strength that confounders would need to have to revert the research conclusions. Specifically, the robustness value $RV_a$ measures the minimum upper bound on both parameters, $R^2_Y\leq RV_a$ and $R^2_D\leq RV_a$, such that the confidence bounds for $\theta_0$ would include a particular value of interest, such as zero, at the significance level $a$.  $RV_a$ thus provide a quick summary of the robustness of an estimated effect --- any confounder with $R^2_Y < RV_a$ and $R^2_D < RV_a$ is logically incapable of explaining away the observed effect.

\begin{figure}[t]
\includegraphics[width=0.85\linewidth]{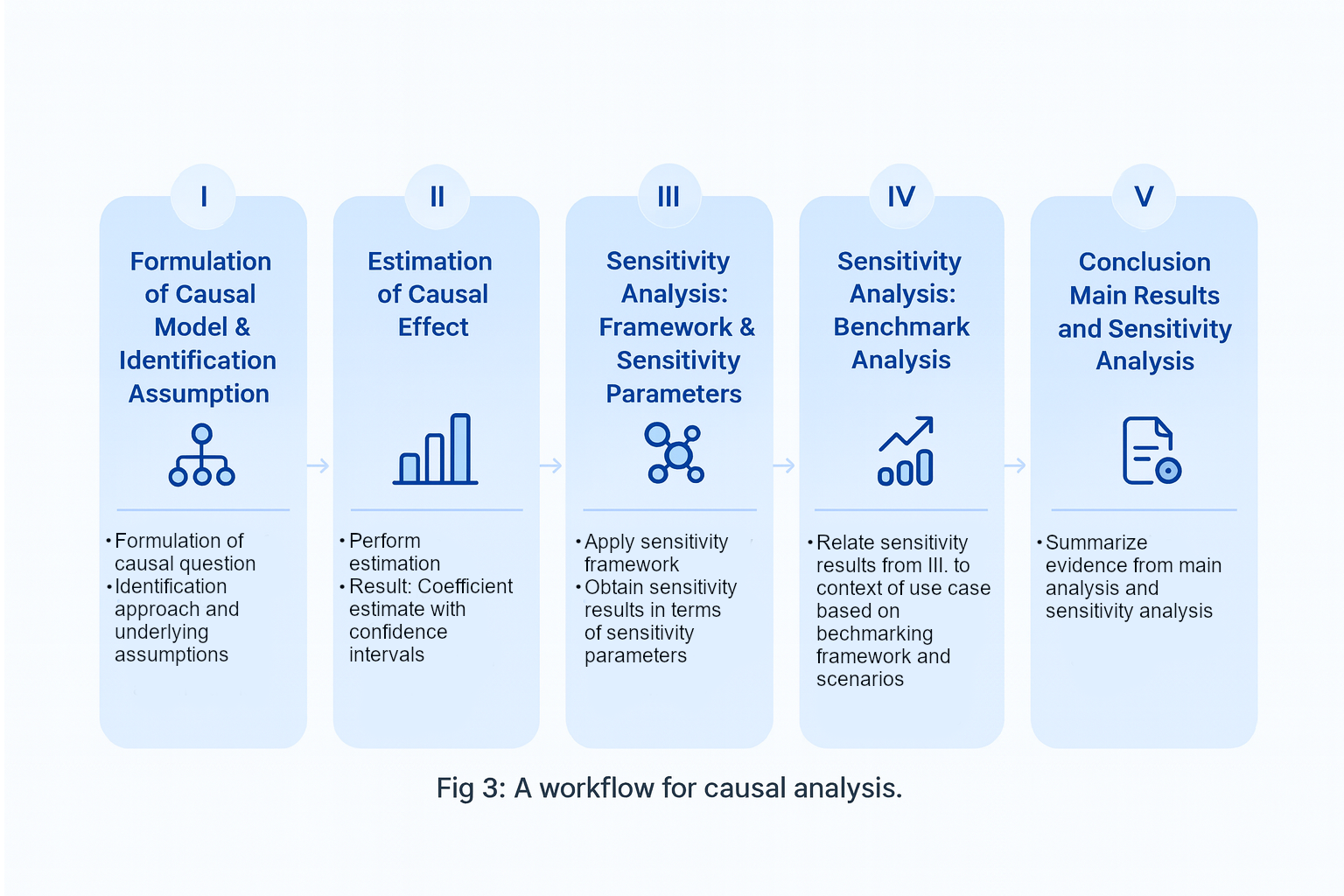}
\caption{A workflow for causal analysis.}
\label{workflow}
\end{figure}

\paragraph{Benchmarking} Sometimes researchers may not have a good idea about the absolute value of the strength of unobserved confounders, but may instead have a notion of their relative importance when compared to key observed covariates. When that is the case, researchers may use the tool of benchmarking \citep{imbens2003sensitivity,cinelli2020making,chernozhukov2023long}, to compute bounds on the strength of unobserved variables if they were as strong or stronger than certain observed covariates.

\subsection{General Sensitivity Analysis}  \label{sec:generalformulation}

Overall, we are interested in the causal parameter $\theta_0$ that can be identified as a linear functional of the long regression function $g_0:=\mathbb{E}[Y|D, X, U]$: 
\begin{align*}
\theta_0 = \mathbb{E}[m(W,g_0)],
\end{align*}
where $m$ is a formula that is affine in $g_0$,
$W$ denotes the full data vector, $W= (Y, D, X, U)$.\footnote{Here, we abstract from the technical details. For a complete presentation, we refer to \citet{chernozhukov2023long}.} 

The key idea of the formulation in \citet{chernozhukov2023long} is to express the long parameter $\theta_0$ as the inner product of the long regression, and weights $\alpha_0$,
\begin{align*}
\theta_0 = \mathbb{E}[m(W, g_0)] = \mathbb{E}[g_0(W)\alpha_0(W)],
\end{align*}
 where $\alpha_0(W)$ is called the Riesz representer of $\theta_0$. The Riesz representer plays a crucial role for causal and debiased ML as it implements a correction for the regularization bias that arises from using ML learners, see \cite{chernozhukov2022automatic}. In many cases, it is possible to obtain an analytical expression for the Riesz representer. In cases where an analytical debiased presentation is not available, an automated procedure can be performed \citep{chernozhukov2022automatic}. 

The expression for the long parameter $\theta_0$ is based on the complete data that includes information on the unobserved confounder(s). To compare the value of the feasible short estimand to the long parameter, we also reformulate $\theta_s$ in terms of its Riesz representation,
 \begin{align*}
\theta_s = \mathbb{E}[m(W_s, g_s)] = \mathbb{E}[g_s(W_s)\alpha_s(W_s)],
\end{align*}
where the subscript $s$ indicates that the quantities refer to the limited available data in the short model, namely, $g_s:= \mathbb{E}[Y|D, X]$ and $W_s = (Y, D, X)$.

\begin{figure}[t]
\includegraphics[width=0.85\linewidth]{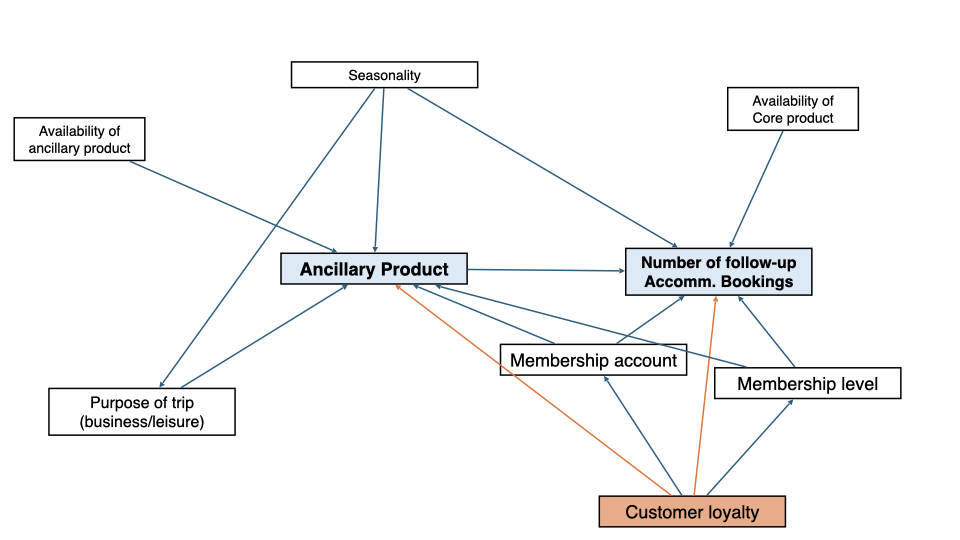}
\caption{Example DAG in the use case.}
\label{dag2}
\end{figure}

\citet{chernozhukov2023long} show that the difference between the long and short parameters is given by the covariance of approximation errors in the regression function and the Riesz representer,
\begin{align*}
\theta_s - \theta_0 = \mathbb{E}\big[(g_s - g_0)(\alpha_s -\alpha_0) \big].
\end{align*}
Furthermore, the squared bias can be further expressed as,
\begin{align*}
\vert \theta_s - \theta_0\vert ^2 = \rho^2 \mathbb{E}\big[ (g_0-g_s)^2\big]\mathbb{E}\big[(\alpha_0-\alpha_s)^2 \big],
\end{align*}
where $\rho^2 \in [0,1]$, is given by the correlation of errors,
\begin{align*}
\rho^2 := \text{Cor}^2(g_0-g_s, \alpha_0-\alpha_s).
\end{align*}

Intuitively, the bias depends on the additional variation that latent variables create in the long regression and the Riesz representer.  This bias formula can be further expressed in terms of   sensitivity parameters  based on $R^2$ measures, which are more directly interpretable.
\begin{align*}
C_Y^2 := \frac{\mathbb{E}(g_0-g_s)^2}{\mathbb{E}(Y-g_s)^2} = R^2_{Y-g_s\sim g_0-g_s}\\
C_D^2:= \frac{\mathbb{E}\alpha_0^2 - \mathbb{E}\alpha_s^2}{\mathbb{E}\alpha_s^2} = \frac{1-R^2_{\alpha_0\sim\alpha_s}}{R^2_{\alpha_0\sim\alpha_s}}.
\end{align*}
In a nonparametric model,  $C_Y^2 = R^2_{Y-g_s\sim g_0-g_s}$ denotes the (nonparametric) partial $R^2$ of the confounders with the outcome. A similar interpretation applies to $1-R^2_{\alpha_0\sim\alpha_s}$, but with respect to the Riesz representer. The term $1-R^2_{\alpha_0\sim\alpha_s}$ measures the share of the residual variation in the long version of the Riesz representer that is explained by the confounding variable(s) $U$.  An appealing feature of these sensitivity parameters is that they are naturally restricted to a range between $0$ and $1$, which improves the interpretability and applicability of the sensitivity framework.

The interpretation of these sensitivity parameters can be further refined depending on the causal parameter of interest, and whether additional parametric assumptions are made. For example, in the setting of prior work by \citet{cinelli2020making},  where the target parameter is a linear projection coefficient, both parameters reduce to traditional linear projection partial $R^2$ measures, $C_Y^2 = R^2_{Y \sim U|D, X}$ and $1-R^2_{\alpha_0\sim\alpha_s} = R^2_{D \sim U|X}$, that is, the partial $R^2$ of $U$ with the treatment $D$ and the outcome $Y$. This partial $R^2$ interpretation is still preserved in a high-dimensional partially linear regression model, cf. \citet{chernozhukov2023long} and \citet{chernozhukov2024applied}. Further examples are presented in the Supplement \ref{example}.

\section{Sensitivity Analysis in a Use Case from Booking.com}\label{sec:usecase}

We now illustrate the application of the previous sensitivity analysis framework in the  use case at Booking.com that was introduced earlier. In this section, we  provide more details on the data set considered and comment on the practical challenges that occurred while applying the omitted variable bias framework of \citet{chernozhukov2023long}. We complement this analysis with a demonstration notebook that leads through the major steps of the analysis based on the implementation with \texttt{DoubleML} for Python \citep{DoubleML2022Python}.\footnote{\url{https://docs.doubleml.org/stable/examples/py_double_ml_sensitivity_booking.html}. More information on the implementation is provided in Section \ref{implementation}.}

The  analysis was based on a pre-analysis plan that defined the sample composition, potential hypotheses, and the definition of the treatment and outcome variable \textit{prior} to the actual estimation step. Customers are defined as visitors of the Booking.com websites and users of the app who have completed an accommodation booking within a time window of six months and have given explicit consent to the usage of their data. The data set is organized in terms of trips, i.e., the purchased ancillary products,  were booked for the same trip as the accommodation booking.  The original data set is relatively large with a total number of observations $>20$ million trips considered. The outcome variable is the number of follow-up bookings of the core product \textit{accommodation} that occurred in a time window of 6 months after the first study period. 

The causal analysis was organized according to the workflow displayed in Figure \ref{workflow}. The first step involves a clear and actionable formulation of the causal question, which is derived from business and management considerations. The primary objective of the evaluation was to determine whether the purchase of ancillary products positively impacts customer relationship, as determined by the ATT. A more precise understanding of the causal model and the potential confounding variables that could threaten identification in the demonstration use case was achieved by brainstorming directed acyclic graphs (DAGs) representing the various causal relationships of interest. Although identifying a single DAG that perfectly fits a specific causal problem can be challenging, the process of modeling and discussing these relationships is crucial for formulating the explicit assumptions necessary to identify the causal effect of interest \citep{cinelli2022crash}. Figure \ref{dag2} shows a DAG that maps the main confounders that should be considered for identification via covariate adjustment.  Note that other approaches to identification, such as instrumental variables, could also be used, but for simplicity, we do not consider them here.

\begin{table}[t]
    \centering
    \begin{tabular}{|c|c|c|c|c|c|}
        \hline
        \multicolumn{6}{|c|}{\textbf{Sensitivity Analysis: Summary}} \\
        \hline
        \multicolumn{3}{|c|}{Significance Level} & \multicolumn{3}{|c|}{level=$0.950$} \\
        \hline
        \multicolumn{3}{|c|}{Sensitivity parameters} & \multicolumn{3}{|c|}{$R^2_Y$=$0.110$; $R^2_D$= $0.003$, $\rho^2$=$1.000$} \\
        \hline
        \multicolumn{6}{|c|}{\textbf{Bounds with CI}} \\
        \hline
        & CI lower & theta lower & theta & theta upper & CI upper \\
        \hline
        d & $0.071$ &  $0.084$ & $0.123$  & $0.162$ & $0.176$ \\
        \hline
        \multicolumn{6}{|c|}{\textbf{Robustness Values}} \\
        \hline
        & $H_0$ & RV (\%) & RVa (\%) & \multicolumn{2}{|c|}{} \\
        \hline
        d & $0.0$ & $5.391$ & $4.816$ &  \multicolumn{2}{|c|}{} \\
        \hline
    \end{tabular}
    \caption{Results from sensitivity analysis based on preferred benchmarking setting.}
    \label{tab:sa_results1}
\end{table}

As the DAG shows, the main confounding variables are \emph{Seasonality}, \emph{Purpose of trip}, \emph{Membership account},  \emph{Membership level}, and \emph{Customer Loyalty}, which is not measured, but partially captured by observed covariates. If the arrows in orange do not exist, then adjusting for the observed confounders would be sufficient for identification of the ATT. On the other hand, if these arrows do exist, this means observed variables are not sufficient to account for customer loyalty, and residual biases may still remain.

Assuming unconfoundedness holds, the second step in the workflow involves estimation, which has been performed using Double Machine Learning through the \texttt{DoubleML} implementation \citep{DoubleML2022Python}. For estimating the nuisance functions, \texttt{LightGBM} \citep{ke2017lightgbm} was employed, with parameter tuning based on cross-validation. We obtained an estimate for the ATT of \(0.123\) with a confidence interval of \([0.107, 0.139]\), indicating a positive, significant, and substantial causal effect for those who purchased an ancillary product. We reiterate that the numerical results presented here do not reflect actual results from the original analysis at Booking.com. The results are based on a simulated data example, which can be replicated using the accompanying notebook.

The previous estimate relies on the assumption of unconfoundedness; specifically, it assumes that the orange arrows in the DAG shown in Figure \ref{dag2} do not exist. The third step of the workflow thus involves performing a sensitivity analysis with the tools of \citet{chernozhukov2023long}.\footnote{We also employed other sensitivity approaches e.g., \citet{vanderweele2011unmeasured}. A comparison is available in the accompanying notebook.}  The results from the sensitivity analysis are presented in Table \ref{tab:sa_results1}. 

In the use case, loyalty-based selection into ancillary product purchases are expected to lead to an overestimation of the ATT. Hence, we focus on the lower bound for the ATT and the corresponding confidence bounds to assess whether the effect estimate would be equal or even smaller than zero or become non-significant in a specific confounding scenario. A confounding scenario is defined in terms of fixed values for the sensitivity parameters $R_Y^2$ and $R_D^2$. These values can either be derived from domain-specific reasoning or being based on benchmarking exercises.  In the absence of a specific scenario, the robustness values provide a concise summary of the robustness of the ATT estimate. For instance, the $RV=5.391\%$ shows that unobserved confounders that explain less than $5.391\%$ of the residual variation of the outcome and of the odds of treatment, are logically incapable of bringing the point estimate of the ATT to zero. If we consider sampling uncertainty, this number reduces to $RV_a=4.816\%$ (at the 5\% significance level).

Table \ref{tab:sa_results1} also shows our preferred confounding scenario, based on benchmarking unobserved confounding against an observed covariate of intermediate strength. In particular, it illustrates how much bias an omitted variable would cause, if it were as strong as ``membership variables.'' These variables were chosen because they are similar in nature to customer loyalty.  This scenario results in calibrated values for the sensitivity parameters of $R_Y^2=0.110$ and $R_D^2=0.003$.  In that scenario, the lower bound of the ATT is $0.084$ and the bound for the lower confidence limit is $0.071$.  This is interpreted as evidence in favor of a positive ATT, which would be robust to unobserved confounding similar the benchmarking variable. Note that in this simulated example, the true value of the ATT is \(0.070\). Thus, in this particular case, the lower limit of the confidence bound provides a better approximation of the true value when compared to a naive estimate that assumes unconfoundedness. In practice, however, the truth is not known. In the actual use case, we also considered stronger, but more implausible, confounding scenarios, benchmarking against all loyalty proxy measures. In the most extreme setting, the ATT was not found to be robust to confounding bias, i.e., the lower bound estimates became negative.

\begin{figure}[t]
\includegraphics[width=\linewidth]{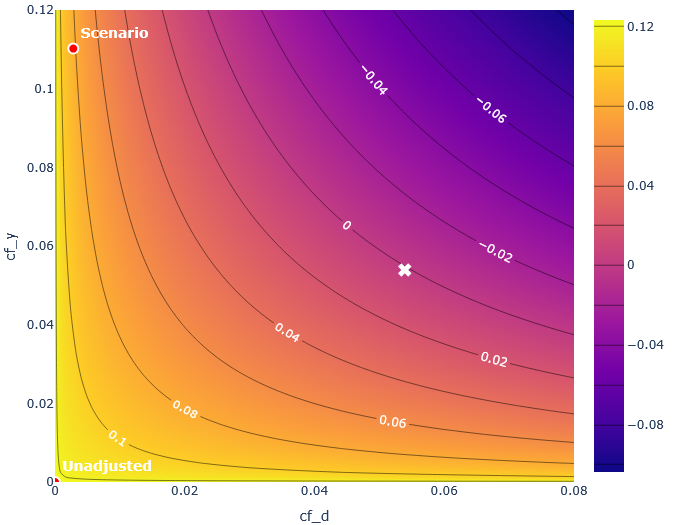}
\caption{Contour plot for sensitivity analysis in the cross-selling use case with benchmarking scenario (red point) and the robustness value (white cross).}
\label{contour2}
\end{figure}

An appealing way to visualize multiple, possibly asymmetric confounding scenarios is through a sensitivity contour plot \citep{imbens2003sensitivity, cinelli2020making, cinelli2022omitted}. In Figure \ref{contour2}, a contour line illustrates all combinations of \(R_Y^2\) and \(R_D^2\) (denoted here as \texttt{cf\_y} and \texttt{cf\_d}) that result in the same lower bias limit of the confidence bound for the causal parameter. The conclusions from sensitivity analysis crucially depends on whether the confounding scenarios are realistic and relevant in a specific use case. Data scientists and domain experts evaluated the plausibility of the different  scenarios. They concluded that benchmarking against all loyalty measures would correspond to an implausibly strong confounding setting. The results of multiple benchmarking exercises of intermediate strengths led to similar results and, hence, to the conclusion of a rather robust and positive ATT.

Finally, integrating sensitivity analysis into the causal analysis workflow has significantly impacted both the analysis and communication of results with business stakeholders. Sensitivity considerations have clarified the fundamental challenges of causal inference with non-experimental data, emphasizing the importance of understanding observational causal analysis and its underlying assumptions. Additionally, sensitivity analysis offered quantitative insights that went beyond intuitive and anecdotal critiques of the unconfoundedness assumption. It provided a clearer understanding of how significant violations of this assumption might be, using domain-specific knowledge relevant to the use case. The sensitivity framework includes practical steps such as calibrating parameters and evaluating benchmark scenarios. By applying sensitivity analysis, data scientists and business stakeholders have gained practical experience that can be applied to future cases.

\section{Conclusion and Outlook}

In this paper, we introduced causal estimation and sensitivity analysis using machine learning techniques through an applied example that mimics a use case of Booking.com. As ML-based causal analysis becomes increasingly prevalent in both academic research and industry applications, integrating sensitivity analysis is essential. It explicitly addresses the risk of violations of identification assumptions, transparently revealing the threats to the validity of observational studies. The framework developed by \citet{chernozhukov2023long} offers a powerful formal approach to managing sensitivity concerns, while being fully compatible with modern Causal Machine Learning. By demonstrating this approach through a practical example, we aim to highlight its utility and relevance, encouraging broader adoption of sensitivity analysis in applied research.





\clearpage

\appendix
\renewcommand{\thesection}{S.\arabic{section}}
\renewcommand{\thesubsection}{S.\arabic{section}.\arabic{subsection}}

\renewcommand{\thefigure}{A.\arabic{figure}}
\counterwithin{figure}{section} 

\section{Supplementary Material}

\subsection{Data Availability}

The data example is fully replicable using the \texttt{DoubleML} library for Python. A detailed example, reproducing the data set and the analysis can be found at \url{https://docs.doubleml.org/stable/examples/py_double_ml_sensitivity_booking.html}.

\subsection{Additional Empirical Results} \label{details}

In the main text, we have not elaborated on the role of the parameter $\rho$ due to space restrictions. The sensitivity parameters $C_D^2$ and $C_Y^2$ serve as measures for the variance in the outcome and treatment variable due to an unobserved confounder. The implications of the confounder are not only depending on those quantities, but also to the degree of how much these variations are related to each other. In the case, where $U$ creates variation in $D$, which, however, is independent from the variation in $Y$ that is induced by $U$, we would have $\rho=0$, i.e., such a setting would not alter the estimate of the causal estimate.  In empirical examples, the value of $\rho$ can be calibrated, for example through benchmarking. In our analyses, we usually started with the setting of $\rho=1$ and reduced the value down to the value that was calibrated using benchmarking.

The contour plot in Figure \ref{contour2} refers to the plug-in estimate of the lower bound of the ATT. It is also possible to construct  similar contour plot for the lower limit of the confidence bound of the ATT.

\subsection{Sensitivity Analysis with the \texttt{DoubleML} library in  Python} \label{implementation}

The Python library \texttt{DoubleML} implements the approach of \cite{chernozhukov2023long} for various causal models, including the nonparametric treatment effect model and a partially linear regression model. Comprehensive documentation and a detailed user guide are available on the project's webpage \url{https://docs.doubleml.org/stable/index.html}.
After running a causal analysis in \texttt{DoubleML}, i.e., after instantiation and fitting of a causal model\footnote{See the workflow at \url{https://docs.doubleml.org/stable/workflow/workflow.html}}, it is possible to perform the sensitivity analysis we discussed here by calling \texttt{sensitivity\_analysis()}, which takes fixed values for the sensitivity parameters \texttt{cf\_d} and \texttt{cf\_y} as input and returns the corresponding bias bounds together with robustness values.\footnote{Note here $\text{cf}_y$ and $\text{cf}_d$ refer to $R^2_Y = C^2_Y$ and $R^2_D=1-R^2_{\alpha \sim \alpha_s}$ parameters.} The values \texttt{theta lower} and \texttt{theta upper} refer to the point estimates of lower bound and upper bound  on the causal parameter (for example the ATT or ATE). The values \texttt{CI lower} and \texttt{CI upper} indicate confidence intervals for the bounds of the causal parameter, thus accounting for sampling uncertainty. The robustness values \texttt{RV} and \texttt{RVa} indicate the minimum upper bound on both sensitivity parameters that would suffice to bring the point estimate (or lower limit of the confidence interval) down to a value of zero.  Contour plots can be generated by calling \texttt{sensitivity\_plot()}. It is possible to manually add confounding scenarios that are then marked by points in the plot. \texttt{sensitivity\_benchmark()} implements the benchmarking procedure by repeating the causal estimation by omitting the candidate benchmark variables. The output provides the values for sensitivity parameters based on comparisons against observed covariates.

\subsection{DML Algorithm} \label{dmlalgo}

The DML algorithm proceeds as follows. Consider a Neyman orthogonal score $\psi(\theta, W; \eta)$, where   $\eta = (g, \alpha)$. Then, given a random sample $(W_i)_{i=1}^N$ of data vectors, (1) randomly partition it into folds $(I_{\ell})_{\ell=1}^L$ of approximately equal size. Denote by $I_{\ell}^c$ the complement of $I_{\ell}$. (2) For each $\ell$, estimate $\widehat \eta_\ell = (\widehat{g}_{\ell}, \widehat{m}_{\ell})$ from observations in $I_{\ell}^c$. (3) Estimate $\theta_0$ as a root of:
$$N^{-1}\sum_{\ell=1}^L\sum_{i\in I_{\ell}} \psi(\theta, W_i; \widehat \eta_\ell)=0.$$

\subsection{Example: Riesz Respresentation for the ATE} \label{example}

In the example of a nonparametric causal model under unconfoundedness, we have that the ATE is defined as
\begin{align}
\theta_0 = \mathbb{E}[Y(1) - Y(0)]=\mathbb{E}[\underbrace{g_0(1,X,U) - g_0(0,X,U)}_{:= m(W, g_0)}].
\end{align}
The long  regression function is
\begin{align*}
g_0(d, X, U) = \mathbb{E}[Y|D=d, X, U],
\end{align*}
with the short version being
\begin{align*}
g_s(d, X) = \mathbb{E}[Y|D=d, X].
\end{align*}
The long Riesz representer for the ATE is
\begin{align*}
\alpha_0(W) = \frac{D}{m_0(X,U)} - \frac{1-D}{1-m_0(X,U)},
\end{align*}
where $m_0(X, U):= \mathbb{E}[D|X, U] = P(D=1|X, U)$ is the long propensity score. Note the RR corresponds to the well known Horvitz-Thompson weights of inverse probability weighting. The short Riesz representer is given by the same weights, excluding $U$, that is,
\begin{align*}
\alpha_s(W_s) = \frac{D}{m_s(X)} - \frac{1-D}{1-m_s(X)},
\end{align*}
where $m_s(X) := \mathbb{E}[D|X] = P(D=1|X)$ refers to the propensity score without the unobserved confounding variables $U$.

The sensitivity parameters for the ATE examples then take the following form 
\begin{align*}
C_Y^2 := \frac{\textrm{Var}(\mathbb{E}[Y|D,X,U]) - \textrm{Var}(\mathbb{E}[Y|D,X])}{\textrm{Var}(Y)-\textrm{Var}(\mathbb{E}[Y|D,X])},
\end{align*}
which equals the non-parametric partial $R^2$ of $Y$ with $U$; and,
\begin{align*}
\tiny
1-R^2_{\alpha \sim \alpha_s}= \frac{\mathbb{E}[1/\mathrm{Var}(D |X,U)]- \mathbb{E}[1/\mathrm{Var}(D|X)]}{\mathbb{E}[1/\mathrm{Var}(D|X, U)]},
\end{align*}
which has an interpretation as the gains in precision in the treatment equation due to $U$.

\subsection{Example: Riesz Representation for the ATT}

Here we provide the Riesz representers for the ATT example discussed in the main text:
\begin{align}
\alpha_0(W) &= \left(\frac{D}{m_0(X, U)} - \frac{1-D}{1-m_0(X, U)}\right) \left(\frac{m_0(X, U)}{p}\right),\nonumber\\
\alpha_s(W_s) &= \left(\frac{D}{m_s(X)} - \frac{1-D}{1-m_s(X)}\right) \left(\frac{m_s(X)}{p}\right),\nonumber
\end{align}
where $p := P(D=1)$. 
The interpretation of $C^2_Y$, $C^2_D$ and $1-R^2_{\alpha \sim \alpha_s}$ are given in the main text.  Note that for simplicity, and to avoid discussing Riesz representation in the main text, the quantity corresponding to $1-R^2_{\alpha \sim \alpha_s}$ of the general formulation is referred simply as~$R^2_D$.


\newpage

\footnotesize
\pagebreak
\bibliographystyle{plainnat}
\bibliography{bibliography}

\begin{thebibliography}{33}
\providecommand{\natexlab}[1]{#1}
\providecommand{\url}[1]{\texttt{#1}}
\expandafter\ifx\csname urlstyle\endcsname\relax
  \providecommand{\doi}[1]{doi: #1}\else
  \providecommand{\doi}{doi: \begingroup \urlstyle{rm}\Url}\fi

\bibitem[Angrist and Pischke(2009)]{angrist2009mostly}
Joshua~D Angrist and J{\"o}rn-Steffen Pischke.
\newblock \emph{Mostly harmless econometrics: An empiricist's companion}.
\newblock Princeton university press, 2009.

\bibitem[Bach et~al.(2022)Bach, Chernozhukov, Kurz, and Spindler]{DoubleML2022Python}
Philipp Bach, Victor Chernozhukov, Malte~S. Kurz, and Martin Spindler.
\newblock {DoubleML} -- {A}n object-oriented implementation of double machine learning in {P}ython.
\newblock \emph{Journal of Machine Learning Research}, 23\penalty0 (53):\penalty0 1--6, 2022.
\newblock URL \url{http://jmlr.org/papers/v23/21-0862.html}.

\bibitem[Bach et~al.(2024{\natexlab{a}})Bach, Kurz, Chernozhukov, Spindler, and Klaassen]{doubleml2024R}
Philipp Bach, Malte~S. Kurz, Victor Chernozhukov, Martin Spindler, and Sven Klaassen.
\newblock {DoubleML}: {A}n object-oriented implementation of double machine learning in {R}.
\newblock \emph{Journal of Statistical Software}, 108\penalty0 (3):\penalty0 1--56, 2024{\natexlab{a}}.
\newblock \doi{10.18637/jss.v108.i03}.
\newblock arXiv:\href{https://arxiv.org/abs/2103.09603}{2103.09603} [stat.ML].

\bibitem[Bach et~al.(2024{\natexlab{b}})Bach, Schacht, Chernozhukov, Klaassen, and Spindler]{bach2024hyperparameter}
Philipp Bach, Oliver Schacht, Victor Chernozhukov, Sven Klaassen, and Martin Spindler.
\newblock Hyperparameter tuning for causal inference with double machine learning: A simulation study, 2024{\natexlab{b}}.

\bibitem[Blackwell(2013)]{blackwell2013selection}
Matthew Blackwell.
\newblock A selection bias approach to sensitivity analysis for causal effects.
\newblock \emph{Political Analysis}, 22\penalty0 (2):\penalty0 169--182, 2013.

\bibitem[Brumback et~al.(2004)Brumback, Hern{\'a}n, Haneuse, and Robins]{brumback2004sensitivity}
Babette~A Brumback, Miguel~A Hern{\'a}n, Sebastien~JPA Haneuse, and James~M Robins.
\newblock Sensitivity analyses for unmeasured confounding assuming a marginal structural model for repeated measures.
\newblock \emph{Statistics in medicine}, 23\penalty0 (5):\penalty0 749--767, 2004.

\bibitem[Carnegie et~al.(2016)Carnegie, Harada, and Hill]{carnegie:jree2016}
Nicole~Bohme Carnegie, Masataka Harada, and Jennifer~L Hill.
\newblock Assessing sensitivity to unmeasured confounding using a simulated potential confounder.
\newblock \emph{Journal of Research on Educational Effectiveness}, 9\penalty0 (3):\penalty0 395--420, 2016.

\bibitem[Chernozhukov et~al.(2018)Chernozhukov, Chetverikov, Demirer, Duflo, Hansen, Newey, and Robins]{chernozhukov2018double}
Victor Chernozhukov, Denis Chetverikov, Mert Demirer, Esther Duflo, Christian Hansen, Whitney Newey, and James Robins.
\newblock Double/debiased machine learning for treatment and structural parameters, 2018.

\bibitem[Chernozhukov et~al.(2022)Chernozhukov, Newey, and Singh]{chernozhukov2022automatic}
Victor Chernozhukov, Whitney~K Newey, and Rahul Singh.
\newblock Automatic debiased machine learning of causal and structural effects.
\newblock \emph{Econometrica}, 90\penalty0 (3):\penalty0 967--1027, 2022.

\bibitem[Chernozhukov et~al.(2023)Chernozhukov, Cinelli, Newey, Sharma, and Syrgkanis]{chernozhukov2023long}
Victor Chernozhukov, Carlos Cinelli, Whitney Newey, Amit Sharma, and Vasilis Syrgkanis.
\newblock Long story short: Omitted variable bias in causal machine learning, 2023.

\bibitem[Chernozhukov et~al.(2024)Chernozhukov, Hansen, Kallus, Spindler, and Syrgkanis]{chernozhukov2024applied}
Victor Chernozhukov, Christian Hansen, Nathan Kallus, Martin Spindler, and Vasilis Syrgkanis.
\newblock Applied causal inference powered by ml and ai, 2024.

\bibitem[Cinelli and Hazlett(2020)]{cinelli2020making}
Carlos Cinelli and Chad Hazlett.
\newblock Making sense of sensitivity: Extending omitted variable bias.
\newblock \emph{Journal of the Royal Statistical Society Series B: Statistical Methodology}, 82\penalty0 (1):\penalty0 39--67, 2020.

\bibitem[Cinelli and Hazlett(2022)]{cinelli2022omitted}
Carlos Cinelli and Chad Hazlett.
\newblock An omitted variable bias framework for sensitivity analysis of instrumental variables.
\newblock \emph{Available at SSRN 4217915}, 2022.

\bibitem[Cinelli et~al.(2019)Cinelli, Kumor, Chen, Pearl, and Bareinboim]{cinelli:icml2019}
Carlos Cinelli, Daniel Kumor, Bryant Chen, Judea Pearl, and Elias Bareinboim.
\newblock Sensitivity analysis of linear structural causal models.
\newblock \emph{International Conference on Machine Learning}, 2019.

\bibitem[Cinelli et~al.(2020)Cinelli, Ferwerda, and Hazlett]{cinelli2020sensemakr}
Carlos Cinelli, Jeremy Ferwerda, and Chad Hazlett.
\newblock sensemakr: Sensitivity analysis tools for ols in r and stata.
\newblock \emph{Available at SSRN 3588978}, 2020.

\bibitem[Cinelli et~al.(2022)Cinelli, Forney, and Pearl]{cinelli2022crash}
Carlos Cinelli, Andrew Forney, and Judea Pearl.
\newblock A crash course in good and bad controls.
\newblock \emph{Sociological Methods \& Research}, page 00491241221099552, 2022.

\bibitem[Cornfield et~al.(1959)Cornfield, Haenszel, Hammond, Lilienfeld, Shimkin, and Wynder]{cornfield1959smoking}
Jerome Cornfield, William Haenszel, E~Cuyler Hammond, Abraham~M Lilienfeld, Michael~B Shimkin, and Ernst~L Wynder.
\newblock Smoking and lung cancer: recent evidence and a discussion of some questions.
\newblock \emph{Journal of the National Cancer institute}, 22\penalty0 (1):\penalty0 173--203, 1959.

\bibitem[Dorie et~al.(2016)Dorie, Harada, Carnegie, and Hill]{dorie2016flexible}
Vincent Dorie, Masataka Harada, Nicole~Bohme Carnegie, and Jennifer Hill.
\newblock A flexible, interpretable framework for assessing sensitivity to unmeasured confounding.
\newblock \emph{Statistics in medicine}, 35\penalty0 (20):\penalty0 3453--3470, 2016.

\bibitem[Frank(2000)]{frank:smr2000}
Kenneth~A Frank.
\newblock Impact of a confounding variable on a regression coefficient.
\newblock \emph{Sociological Methods \& Research}, 29\penalty0 (2):\penalty0 147--194, 2000.

\bibitem[Frank et~al.(2008)Frank, Sykes, Anagnostopoulos, Cannata, Chard, Krause, and McCrory]{frank:eepa2008}
Kenneth~A Frank, Gary Sykes, Dorothea Anagnostopoulos, Marisa Cannata, Linda Chard, Ann Krause, and Raven McCrory.
\newblock Does nbpts certification affect the number of colleagues a teacher helps with instructional matters?
\newblock \emph{Educational Evaluation and Policy Analysis}, 30\penalty0 (1):\penalty0 3--30, 2008.

\bibitem[Frank et~al.(2013)Frank, Maroulis, Duong, and Kelcey]{frank2013would}
Kenneth~A Frank, Spiro~J Maroulis, Minh~Q Duong, and Benjamin~M Kelcey.
\newblock What would it take to change an inference? {U}sing {R}ubin's causal model to interpret the robustness of causal inferences.
\newblock \emph{Educational Evaluation and Policy Analysis}, 35\penalty0 (4):\penalty0 437--460, 2013.

\bibitem[Hosman et~al.(2010)Hosman, Hansen, and Holland]{hosman2010sensitivity}
Carrie~A Hosman, Ben~B Hansen, and Paul~W Holland.
\newblock The sensitivity of linear regression coefficients' confidence limits to the omission of a confounder.
\newblock \emph{The Annals of Applied Statistics}, pages 849--870, 2010.

\bibitem[Imai et~al.(2010)Imai, Keele, Yamamoto, et~al.]{imai2010identification}
Kosuke Imai, Luke Keele, Teppei Yamamoto, et~al.
\newblock Identification, inference and sensitivity analysis for causal mediation effects.
\newblock \emph{Statistical science}, 25\penalty0 (1):\penalty0 51--71, 2010.

\bibitem[Imbens(2003)]{imbens2003sensitivity}
Guido~W Imbens.
\newblock Sensitivity to exogeneity assumptions in program evaluation.
\newblock \emph{The American Economic Review}, 93\penalty0 (2):\penalty0 126--132, 2003.

\bibitem[Ke et~al.(2017)Ke, Meng, Finley, Wang, Chen, Ma, Ye, and Liu]{ke2017lightgbm}
Guolin Ke, Qi~Meng, Thomas Finley, Taifeng Wang, Wei Chen, Weidong Ma, Qiwei Ye, and Tie-Yan Liu.
\newblock Lightgbm: A highly efficient gradient boosting decision tree.
\newblock \emph{Advances in neural information processing systems}, 30, 2017.

\bibitem[Middleton et~al.(2016)Middleton, Scott, Diakow, and Hill]{middleton2016bias}
Joel~A Middleton, Marc~A Scott, Ronli Diakow, and Jennifer~L Hill.
\newblock Bias amplification and bias unmasking.
\newblock \emph{Political Analysis}, 24\penalty0 (3):\penalty0 307--323, 2016.

\bibitem[Oster(2017)]{oster:jbes2017}
Emily Oster.
\newblock Unobservable selection and coefficient stability: Theory and evidence.
\newblock \emph{Journal of Business \& Economic Statistics}, pages 1--18, 2017.

\bibitem[Robins(1999)]{robins1999association}
James~M Robins.
\newblock Association, causation, and marginal structural models.
\newblock \emph{Synthese}, 121\penalty0 (1):\penalty0 151--179, 1999.

\bibitem[Rosenbaum(2002)]{rosenbaum2002gamma}
Paul~R Rosenbaum.
\newblock Observational studies.
\newblock In \emph{Observational studies}, pages 1--17. Springer, 2002.

\bibitem[Rosenbaum and Rubin(1983)]{rosenbaum1983assessing}
Paul~R Rosenbaum and Donald~B Rubin.
\newblock Assessing sensitivity to an unobserved binary covariate in an observational study with binary outcome.
\newblock \emph{Journal of the Royal Statistical Society. Series B (Methodological)}, pages 212--218, 1983.

\bibitem[Vanderweele and Arah(2011)]{arah2011}
Tyler~J. Vanderweele and Onyebuchi~A. Arah.
\newblock {Bias formulas for sensitivity analysis of unmeasured confounding for general outcomes, treatments, and confounders.}
\newblock \emph{Epidemiology (Cambridge, Mass.)}, 22\penalty0 (1):\penalty0 42--52, January 2011.
\newblock ISSN 1531-5487.
\newblock \doi{10.1097/ede.0b013e3181f74493}.
\newblock URL \url{http://dx.doi.org/10.1097/ede.0b013e3181f74493}.

\bibitem[VanderWeele and Arah(2011)]{vanderweele2011unmeasured}
Tyler~J VanderWeele and Onyebuchi~A Arah.
\newblock Unmeasured confounding for general outcomes, treatments, and confounders: bias formulas for sensitivity analysis.
\newblock \emph{Epidemiology (Cambridge, Mass.)}, 22\penalty0 (1):\penalty0 42, 2011.

\bibitem[Zhang et~al.(2021)Zhang, Cinelli, Chen, and Pearl]{zhang2021exploiting}
Chi Zhang, Carlos Cinelli, Bryant Chen, and Judea Pearl.
\newblock Exploiting equality constraints in causal inference.
\newblock In \emph{International Conference on Artificial Intelligence and Statistics}, pages 1630--1638. PMLR, 2021.

\end{thebibliography}

\end{document}